\documentclass[aip, long, numerical bibliography, (default),jmp, reprint, superscriptaddress, nofootinbib]{revtex4-1}
\pdfoutput=1
\usepackage{graphicx}
\usepackage{dcolumn}
\usepackage{bm}
\usepackage{epstopdf}
\usepackage{xr}
\usepackage{amsmath}
\usepackage{color}
\usepackage[normalem]{ulem}
\usepackage{wasysym}

\begin{document}
\markboth{\jobname}{\jobname .tex}

\title{Strong electronic interaction and multiple quantum Hall ferromagnetic phases  in trilayer graphene}
\author{Biswajit Datta}
\affiliation{Department of Condensed Matter Physics and Materials Science, Tata Institute of Fundamental Research, Homi Bhabha Road, Mumbai 400005, India}
\author{Santanu Dey}
\affiliation{Department of Astronomy and Astrophysics, Tata Institute of Fundamental Research, Homi Bhabha Road, Mumbai 400005, India}
\author{Abhisek Samanta}
\affiliation{Department of Theoretical Physics, Tata Institute of Fundamental Research, Homi Bhabha Road, Mumbai 400005, India}
\author{Abhinandan Borah}
\affiliation{Department of Condensed Matter Physics and Materials Science, Tata Institute of Fundamental Research, Homi Bhabha Road, Mumbai 400005, India}
\author{Kenji Watanabe}
\affiliation{Advanced Materials Laboratory, National Institute for Materials Science, 1-1 Namiki, Tsukuba 305-0044, Japan}
\author{Takashi Taniguchi}
\affiliation{Advanced Materials Laboratory, National Institute for Materials Science, 1-1 Namiki, Tsukuba 305-0044, Japan}
\author{Rajdeep Sensarma}
\homepage{sensarma@theory.tifr.res.in }
\affiliation{Department of Theoretical Physics, Tata Institute of Fundamental Research, Homi Bhabha Road, Mumbai 400005, India}
\author{Mandar M. Deshmukh}
\homepage{deshmukh@tifr.res.in}
\affiliation{Department of Condensed Matter Physics and Materials Science, Tata Institute of Fundamental Research, Homi Bhabha Road, Mumbai 400005, India}

\begin{abstract}

\end{abstract}
\maketitle

There is an increasing interest in the electronic properties of few layer graphene~\cite{yacoby_graphene:_2011,taychatanapat_quantum_2011,bao_stacking-dependent_2011,kumar_integer_2011,zhang2012hund,henriksen_quantum_2012,craciun2009trilayer,campos_landau_2016,lee2013broken,stepanov_tunable_2016} as it offers a platform to study electronic interactions because the dispersion of bands can be tuned with number and stacking of layers~\cite{yacoby_graphene:_2011} in combination with electric field~\cite{serbyn_new_2013}. However, electronic interaction becomes important only in very clean  devices  and so far the trilayer graphene experiments are understood within non-interacting electron picture.  Here, we report evidence of strong electronic interactions and quantum Hall ferromagnetism (QHF) seen in  ABA trilayer graphene (ABA-TLG). Due to high mobility $\sim$500,000 cm$^2$V$^{-1}$s$^{-1}$  in our device compared to previous studies, we find all symmetry broken states and that Landau Level (LL) gaps are enhanced by interactions; an aspect explained by our self-consistent Hartree-Fock (H-F) calculations. Moreover, we observe  hysteresis as a function of filling factor ($\nu$) and spikes in the longitudinal resistance~\cite{poortere_resistance_2000,jungwirth_resistance_2001} which, together,  signal the formation of QHF states at low magnetic field.





Mesoscopic experiments tuning the relative importance  of electronic interactions to observe complex ordered phases have a rich past~\cite{girvin_spin_2007}. While one class of experiments were conducted on bilayer two dimensional electron systems (2DES) realized in semiconductor heterostructures, the other class of experiments focussed on probing  multiple interacting sub-bands in quantum well structures~\cite{eom_quantum_2000}. ABA-TLG also provides a natural platform to observe such multi-subband physics. The presence of the multiple bands and their Dirac nature lead to the possibility of observing interesting interplay of electronic interactions in different channels leading to novel phases of the quantum Hall state.

\begin{figure}
\includegraphics[width=7.0cm]{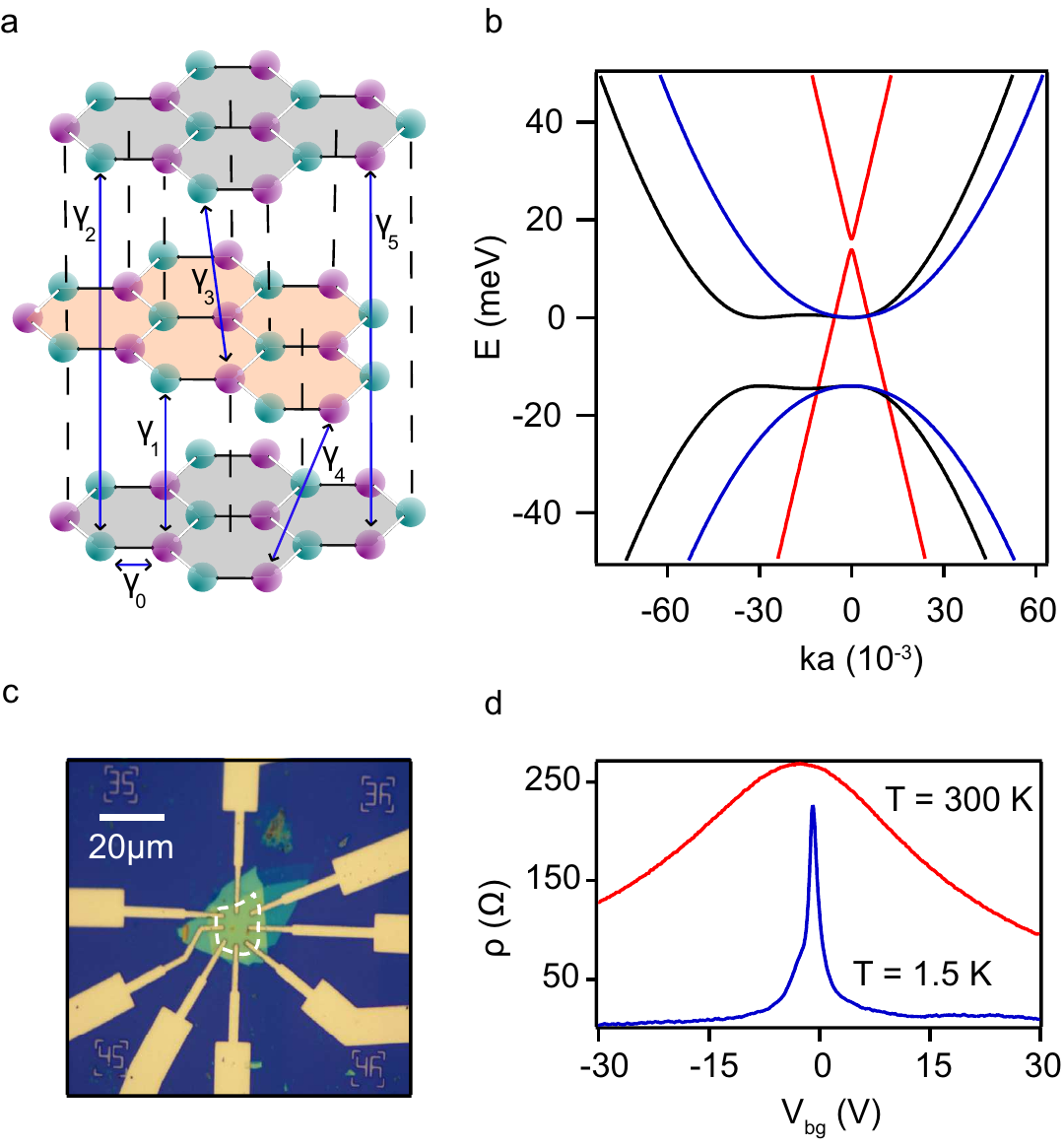}
\caption{ \label{fig:fig1} \textbf{ABA-TLG device.} a) Schematic of the crystal structure of ABA-TLG with all hopping parameters.  b) Low energy band structure of ABA-TLG around k$ _{-} $ point (-$\frac{4\pi}{3}$,0) in the Brillouin zone. The wave vector is normalised with the inverse of the lattice constant (a = 2.46 $\AA$) of graphene. Black and blue lines denote the BL bands along k$ _{x} $ and k$ _{y}$ direction in the Brillouin zone whereas the red line denotes the ML band along both k$ _{x} $ and k$ _{y} $.  ML bands are separated by $\sim\delta + \frac{\gamma_2}{2} - \frac{\gamma_5}{2}$ = 2 meV and BL bands are separated by $\sim\frac{|\gamma_2|}{2}$ = 14 meV. However, there is no band gap in total, semi-metallic nature of ABA-TLG is clear from the band overlap. c) Optical image of the hBN encapsulated trilayer graphene device; white dashed line indicates the boundary of the ABA-TLG. Substrate consists of 30 nm thick hBN and 300 nm thick Silicon dioxide (SiO$_2$) coated highly p doped Silicon which also serves as global back gate. d) Room temperature and low temperature four probe  resistivity  of the device as a function of V$_{bg}$. }
\end{figure}

Fig.~\ref{fig:fig1}a shows the lattice structure of ABA-TLG  with all the hopping parameters.  We use   Slonczewski-Weiss-McClure (SWMcC) parametrization of the tight binding model for ABA-TLG~\cite{mccann_landau-level_2006,koshino_parity_2010} (with hopping parameters $\gamma_0,\gamma_1, \gamma_2, \gamma_5$ and $\delta$) to calculate its low energy bandstructure. Definitions of all the hopping parameters are evident from Fig.~\ref{fig:fig1}a and $\delta$ is the onsite energy difference of two inequivalent carbon atoms on the same layer. Its band structure, shown in Fig.~\ref{fig:fig1}b, consists of both  monolayer-like (ML) linear and bilayer-like (BL) quadratic bands~\cite{serbyn_new_2013,morimoto_gate-induced_2013}.

 Fig.~\ref{fig:fig1}c shows an optical image of the  device where the ABA-TLG graphene is encapsulated between two hexagonal boron nitride (hBN) flakes~\cite{wang_one-dimensional_2013}. Four  probe resistivity ($\rho$) of the device is shown in Fig.~\ref{fig:fig1}d.  The low disorder in the device is reflected in  high mobility $\sim$500,000 cm$^2$V$^{-1}$s$^{-1}$ on electron side and $\sim$800,000 cm$^2$V$^{-1}$s$^{-1}$ on hole side; this leads to carrier mean free path in excess of $7\mu$m (Supplementary Information).


\begin{figure}
\includegraphics[width=7.5cm]{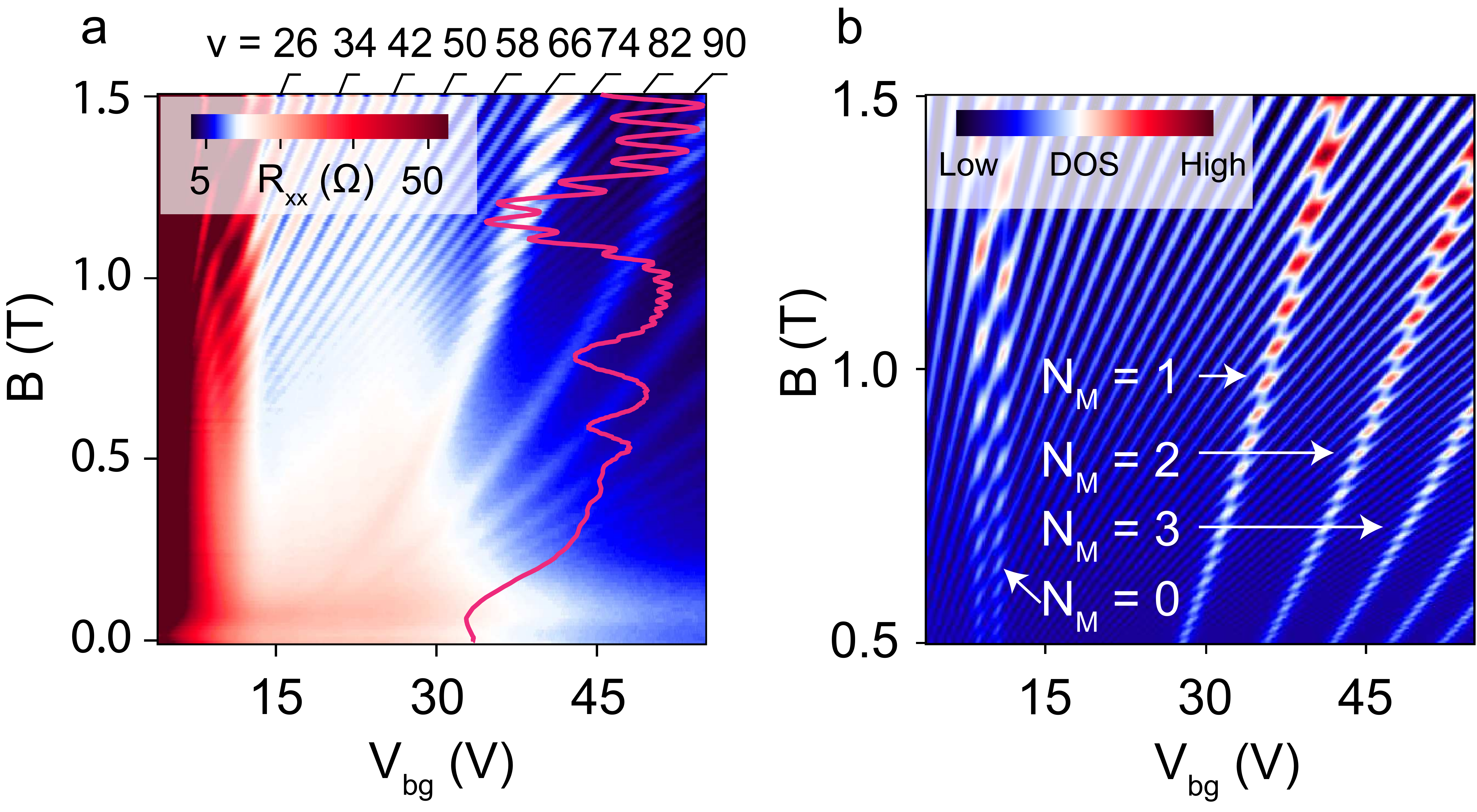}
\caption{ \label{fig:fig2} \textbf{Low magnetic field fan diagram.} a) Colour plot of $R_{xx}$ as a function of V$_{bg}$ and B up to 1.5~T. The LL crossings arising from ML band and BL bands are clearly seen. Each parabola is formed by the repetitive crossings of a particular ML LL with other BL LLs. Crossing between any two LLs  show up as $R_{xx} $ maxima in transport measurement due to high DOS  at the crossing points. The overlaid pink line shows a line slice at  V$_{bg}$ = 50~V.
b) DOS corresponding to Fig.~\ref{fig:fig2}a. N$_M$ = 1 labelled parabola refer to  all the crossing points arising from the crossings of  N$_M$ = 1 LL with other BL LLs. Other labels have similar meaning. N$_M$ = 0 LL does not disperse with B, hence the crossings form a straight line parallel to B axis. Minimum B is taken as 0.5~T to keep  finite number of LLs in the calculation. Horizontal axis is converted from charge density to an equivalent V$_{bg}$  after normalizing it by the capacitance per unit area (C$_{bg}$) for the ease of comparison with experimental fan diagram. C$_{bg}$ is determined from the high B quantum Hall data which matches well with the geometrical capacitance per unit area of 30 nm hBN and 300 nm SiO$_2$: C$_{bg}\sim$105 $\mu F$ m$^{-2}$.}
\end{figure}


We next consider the magnetotransport in ABA-TLG that reveals  the presence of LLs arising from both ML and BL bands. The LLs are characterized by the following quantum numbers: (i) $N_M$ ($N_B$) defines the LL index with M (B) indicating monolayer (bilayer)-like LLs (ii) + (-) denotes the valley index of the LLs (iii) $\uparrow$ ($ \downarrow $) denotes the spin quantum number of the electrons. All the data shown in this paper, are taken at 1.5~K.   Fig.~\ref{fig:fig2}a shows the measured longitudinal resistance (R$_{xx}$) as a function of gate voltage (V$_{bg}$) and magnetic field (B) in the low B regime (see Supplementary Information for more data). Observation of LLs up to very high filling factor $\nu$ = 118 confirms the high quality of the device. Along with the usual straight lines in the fan diagram, we find additional interesting parabolic lines which arise because of LL crossings. Fig.~\ref{fig:fig2}b shows the  calculated non-interacting density of states (DOS)  in the same parameter range which matches very well with the measured resistance.  We find that the low B data can be well understood in terms of non-interacting picture and it allows determination of the band parameters.

\begin{figure*}
\includegraphics[width=15.0cm]{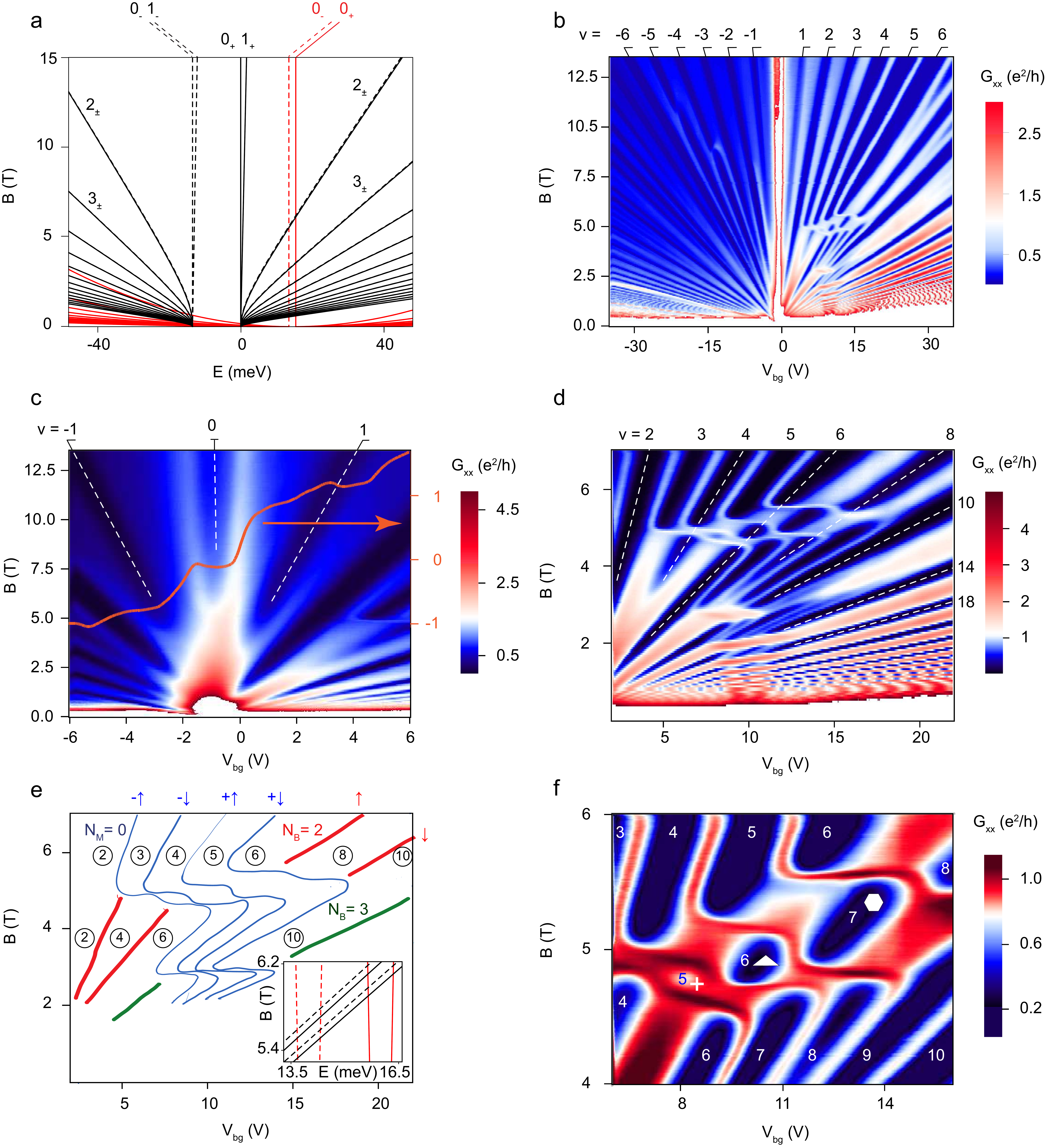}
\caption{ \label{fig:fig3} \textbf{Landau level crossings  and resulting quantum Hall ferromagnetic ground states.} a) Calculated low energy spectra using SWMcC parametrization of the tight binding model for ABA-TLG~\cite{mccann_landau-level_2006,koshino_parity_2010,serbyn_new_2013}. Red and black lines denote the ML and BL LLs respectively. Solid and dashed lines denote LLs coming from k$ _{+} $ and k$ _{-} $  valleys respectively. Labelled numbers represent the LL indices of the corresponding LLs.
b) Colour scale plot of  G$_{xx}$, showing the LL fan diagram. The filling factors measured independently from the $G_{xy}$ are labelled in every plot. As a function of the B  one can observe several crossings on electron and hole side. The data shown in Fig.~\ref{fig:fig2}a forms a very thin slice of the low B data shown in this panel.
c) Zoomed in fan diagram around charge neutrality point showing occurrence of $\nu$ = 0 from $\sim$6~T:  G$_{xx}$ shows a dip and G$_{xy}$ shows a plateau at $\nu$ = 0. The overlaid red line presents  G$_{xy}$ at 13.5 T which shows the occurrence of $\nu$ = -1,0 and 1 plateaus.
d) Zoomed-in recurrent crossings of $N_{M}= 0$ LL with different BL LLs.
e) The lines indicate the LLs seen in the data shown in Fig.~\ref{fig:fig3}d and their crossings. Circled numbers denote the filling factors.
f) A further zoomed in view of the parameter space showing LL crossing of fourfold symmetry broken $N_{M}$ = 0 LL with spin split $N_{B}$ = 2 LL.}

\end{figure*}

We now consider the LL fan diagram for a larger range of V$_{bg}$ and B. Fig.~\ref{fig:fig3}a shows the calculated~\cite{koshino_parity_2010,mccann_landau-level_2006,serbyn_new_2013} energy dispersion of the spin degenerate LLs with B. We use zero electric field approximation (Supplementary Information) which is reasonable and consistent with experiment. All the band parameters of multilayer graphene are not known precisely, so, we refine the band parameters  a little over the known values for bulk graphite\cite{dresselhaus_intercalation_2002} in order to understand our experimental data. We find $\gamma_0$ = 3.1eV,$\gamma_1$ = 0.39eV, $\gamma_2$ = -0.028eV,   $\gamma_5$ = 0.01eV and  $\delta$ = 0.021eV  best describe our data. Fig.~\ref{fig:fig3}b shows the main fan diagram where the measured longitudinal conductance (G$_{xx}$) is plotted as a function of V$_{bg}$ and B. Due to lack of inversion symmetry, valley degeneracy is not protected in ABA-TLG, it breaks up with increasing B and reveals all the symmetry broken filling factors as seen in Fig.~\ref{fig:fig3}b.



Fig.~\ref{fig:fig3}c shows measured G$_{xx}$ focusing on the $\nu$ = 0 state,  which shows a dip right at the charge neutrality point, evident for  B  $>$ 6~T.  Corresponding transverse conductance (G$_{xy}$) shows a plateau at zero  indicating the occurrence of  the $\nu$ = 0 state. While, the $\nu$ = 0 plateau has been observed in monolayer graphene~\cite{zhang_landau-level_2006} and in bilayer graphene~\cite{zhao_symmetry_2010} (for B more than $ \sim$15~T - 25~T),  this is the first observation of $\nu$ = 0 state in trilayer graphene at such a low B. A marked reduction in disorder allows observation of the $\nu$ = 0 state in our device.

Focusing on the electron side, Fig.~\ref{fig:fig3}d and \ref{fig:fig3}e show the experimentally measured LL fan diagram and labelled LLs, respectively. We see that the presence of $N_M = 0$ LL gives rise to a series of vertical crossings along the B axis as is expected from the LL energy diagram (Fig.~\ref{fig:fig3}a).   The highest crossing along the B axis appears when $N_M = 0$ crosses with $N_B = 2$  LL at $\sim5$~T.

From the complex fan diagram, seen in Fig.~\ref{fig:fig3}d and \ref{fig:fig3}e, we can see both above and below the topmost LL crossing ($V_{bg}\sim$10~V and $B\sim$5~T),  $N_M = 0$ LL is completely symmetry broken and $N_B = 2$ LL quartet on the other hand  becomes  two fold split at $\sim$3.5~T. The crossing between N$_M$ = 0 and N$_B$ = 2 LLs gives rise to three ring-like structures.  Calculated LL energy spectra near the topmost crossing (Fig.~\ref{fig:fig3}e inset) shows that spin splitting is larger than valley splitting for $N_B = 2$ LL but valley splitting dominates over spin splitting for $N_M = 0$ LL. We note that valley splitting of $N_M$ = 0 is very large compared to other ML LLs; which arises because ML bands are  gapped in ABA-TLG  unlike in monolayer  graphene.  As one follows the $N_M = 0$ LL down towards B = 0 one observes successive LL crossings of $N_M = 0$ with $N_B = $ 2,3,4 ..... The sharp abrupt bends in the fan diagram occur due to the change of the order of filling up of LLs after crossings and the fact that the horizontal axis is charge density (proportional to $V_{bg}$ and not LL energy). When these crossings are extrapolated to $B = 0$, we see that $N_M = 0$ LL is valley split as expected  from the LL energy diagram Fig.~\ref{fig:fig3}a.




We next discuss experimental signatures that point towards the importance of interaction. Observation of spin split N$_{M}$ = 0 LL at B $\sim$2~T cannot be explained from the non-interacting Zeeman splitting for $\Gamma  \sim$1.5~meV on electron side, estimated from the Dingle plot. Also, large ratio of
transport scattering time ($\tau_t$) to quantum scattering time ($\tau_q$)($\frac{\tau_t}{\tau_q} \approx$ 49) indicates that small angle scattering is  dominant, a signature of the long range nature of the Coulomb  potential~\cite{Das_Sarma_transport_scatter_2008,coleridge_small-angle_1991,Knap_interaction_in_SdH} (Supplementary Information). We also measure activation gap  for the symmetry broken states $\nu = 2, 3, 4, 5, 7$ at B = 13.5~T, and find significantly higher gaps than the non-interacting spin-splitting. For $\nu = 3$ and 5, Fermi energy (E$_F$) lies in spin polarized gap of N$_M$ = 0 LL in K$_-$ and K$_+$ valley respectively. Measured energy gap at $\nu = 3$  is        $\sim$5.1~meV and  at $\nu = 5$  is $\sim$2.8~meV whereas free electron Zeeman splitting is $\sim$1.56~meV at B = 13.5~T (Supplementary Information). We note that typically the transport gap tends to underestimate the real gap due to the LL broadening, so actual single particle gap might be even larger.

Interaction results in symmetry broken states at low B that are QHF states. For the data in Fig.~\ref{fig:fig3}d, $\nu$  =  2,3,4,5 are QHF states  for  $B>5.5$~T. Similarly, $\nu = $ 7,8,9  are also QHF states for 5.5~T~$>B>$~4~T. In fact the LLs associated with $\nu = $ 3,4,5 after crossing are the same ML LLs which are responsible for $\nu=$ 7,8,9 before crossing (Fig.~\ref{fig:fig3}e). The crossings result in three ring-like structures marked by  $+$, $\bigtriangleup$ and $\hexagon$ in Fig.~\ref{fig:fig3}f.

\begin{figure}
\includegraphics[width=8cm]{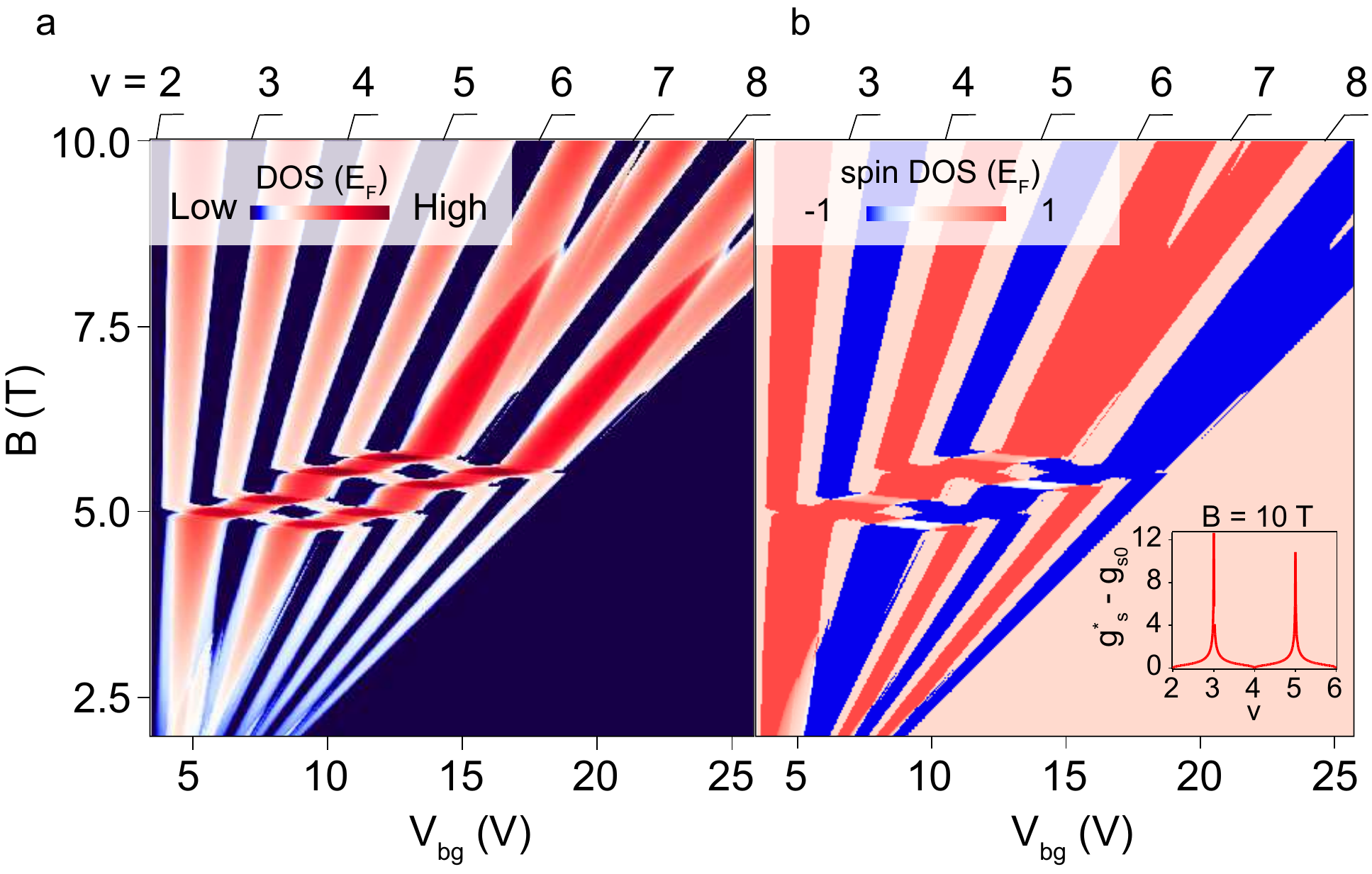}
\caption{ \label{fig:fig4}\textbf{Theoretical calculation of DOS and spin-polarization.} a) DOS at the Fermi level as a function of  V$_{bg}$ and B. This matches the fan diagram seen in the experiment.  b) The magnetization in the system as a function of V$_{bg}$ and $B$, where density is converted to an equivalent V$_{bg}$, described in Fig.~\ref{fig:fig2}b caption. The LL crossing regions clearly show presence of spin polarization in the system. Inset shows calculated enhanced spin {\it g}-factor above the bare value 2 for N$_M$ = 0 spin and valley split LLs.}
\end{figure}

Now we discuss theoretical calculations to show that electronic interactions are crucial in obtaining a quantitative understanding of the experimental data. The theoretical calculations focus on the $N_M=0$ and $N_B=2$  LLs, which form the most prominent LL crossing pattern in our data. The effect of disorder is incorporated within a self-consistent Born approximation (SCBA)~\cite{Ando1,Ando2}, while electronic interactions are included by considering the exchange corrections to the LL spectrum due to a statically screened Coulomb interaction~\cite{Ando3,Gorbar1} in a self-consistent way.  Fig.~\ref{fig:fig4}a shows the DOS at E$_F$ as a function of V$_{bg}$ and $B$, which matches with the experimental results on the G$_{xx}$.

Our calculations also provide insight about the polarization of the states inside the ring-like structures (Fig.~\ref{fig:fig3}f). We find that although the filling factor of region $\bigtriangleup$  is the same as that of regions $\nu$ = 6 above and below,  electronic configurations of these states are different. Fig.~\ref{fig:fig4}b  shows the spin-resolved DOS at E$_F$ as a function of  V$_{bg}$ and $B$. We find total spin polarization (integrated spin DOS) in region $\bigtriangleup$ is non zero (Supplementary Information), but it vanishes in regions $\nu$ = 6 above and below the ring structure.
Fig.~\ref{fig:fig4}b inset shows the calculated exchange enhanced spin {\it g}-factors. This shows significant increase over bare value of {\it g} in the spin polarized states -- in agreement with the large gap observed at $\nu$ = 3 and 5 in experiment.

\begin{figure}
\includegraphics[width=8cm]{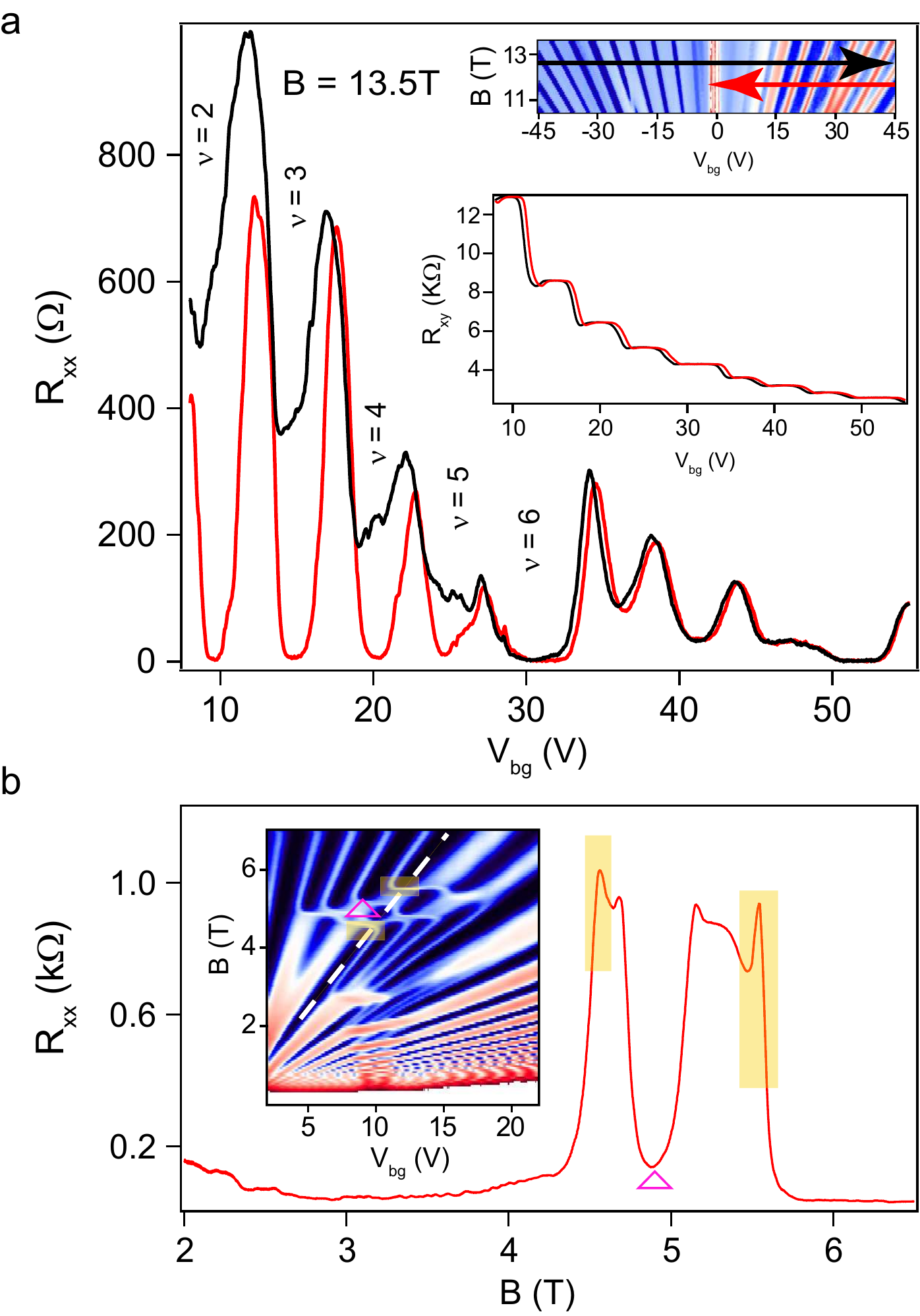}
\caption{ \label{fig:fig5}\textbf{Hysteresis in the longitudinal resistance as a function of the filling factor and observation of resistance spikes.} a) Measurement of $R_{xx}$ as a function of V$_{bg}$ at B = 13.5~T  in the two directions as shown in (top inset) the measurement parameter space. Largest hysteresis is seen for the spin and valley polarized  $N_M = 0$ LL. The lower inset shows simultaneously measured $R_{xy}$ that exhibits clear quantization plateaus in the two sweep directions. b) $R_{xx}$ plotted along the dashed line shown in the parameter space. Spikes in resistance, shaded in yellow, correspond to boundaries of the region marked $\bigtriangleup$.}
\end{figure}

The key role of interactions is also reflected  in the hysteresis of $R_{xx}$ in the vicinity of the symmetry broken QHF states. QHF has been studied in 2DES using semiconductors \cite{eom_quantum_2000,poortere_resistance_2000} and in monolayer graphene \cite{nomura_quantum_2006,young_spin_2012}. We vary filling factor by changing V$_{bg}$ at a fixed B (Fig.~\ref{fig:fig5}a) and observe that the sweep up and down of V$_{bg}$ shows a hysteresis in $R_{xx}$ which can be attributed to the occurrence of pseudospin magnetic order at the symmetry broken filling factors~\cite{piazza_first-order_1999} (see Supplementary Information for hysteresis data at $\nu$ = 7,8,9 at 3.5~T). Corresponding hysteresis is absent in simultaneously measured $R_{xy}$  (Fig.~\ref{fig:fig5}a inset). Hysteresis in $R_{xx}$ with V$_{bg}$ is also absent without B (Supplementary Information). The pinning, that causes the hysteresis could be due to residual disorder within the system as the domains of the QHF evolve. Transport measurements show appearance of   $R_{xx}$ spikes  around the crossing of N$_M$ = 0 and N$_B$ = 2 LLs (Fig.~\ref{fig:fig5}b).  One possible explanation of the spike in R$_{xx}$~\cite{muraki_charge_2001} is the edge state transport along domain wall boundaries as studied earlier in semiconductors~\cite{poortere_resistance_2000,jungwirth_resistance_2001}.

In summary, we see interaction plays an important role to enhance the {\it g}-factor and favours the formation of QHF states at low B and at relatively higher temperature.  ABA-TLG is the simplest system that has both massless and massive Dirac fermions, giving rise to an intricate and rich pattern of LLs that,  through their crossings,  can allow a detailed study of the effect of interaction at sufficiently low temperature. The ability to image these QHF states using modern scanning probe techniques at low magnetic fields could provide insight into these states that have never been imaged previously.  In future, experiments on multilayer graphene, exchange coupled with a ferromagnetic insulating substrate \cite{wang_proximity_ferro_2015}, can lead to the possibility of observing  interesting interplay of  QHF with the proximity induced ferromagnetic order.




\subsection*{Acknowledgements}
We thank Allan MacDonald,  Jainendra Jain, Jim Eisenstein, Fengcheng Wu, Vibhor Singh, Shamashis Sengupta, Chandni U.  for discussions and comments on the manuscript. We also thank John Mathew, Sameer Grover and Vishakha Gupta for experimental assistance. We acknowledge Swarnajayanthi Fellowship of Department of Science and Technology (for MMD) and Department of Atomic Energy of Government of India for support. Preparation of hBN single crystals are supported by the Elemental Strategy Initiative conducted by the MEXT, Japan and a Grant-in-Aid for Scientific Research on Innovative Areas ``Science of Atomic Layers" from JSPS.

\subsection*{Author Contributions}
B.D. fabricated the device, conceived the experiments and analysed the data.   M.M.D., A.B. and B.D. contributed to development of the device fabrication process.  K. W and T. T. grew the hBN crystals.  S.D., A.S. and B.D. did the calculations under the supervision of R.S. B.D. and M.M.D co-wrote the manuscript,  R.S. provided input on the manuscript. All authors commented on the manuscript. M.M.D supervised the project.

\bibliography{trilayer}

\end{document}